\documentclass[a4paper,fleqn]{cas-dc}

\usepackage[authoryear]{natbib}

\usepackage{amsmath}
\usepackage{amssymb}
\usepackage{subcaption, multirow, array}
\usepackage{booktabs}
\usepackage{xspace}
\usepackage{mathtools}
\usepackage{siunitx}
\usepackage{xcolor}

\mathchardef\mhyphen="2D
\newcommand{\specialcell}[2][l]{%
  \begin{tabular}[#1]{@{}l@{}}#2\end{tabular}}
\newcommand{\method}{CAVE}
\newcommand{\subpara}[1]{\vspace{5pt} \noindent \textbf{#1.}}

\makeatletter
\DeclareRobustCommand\onedot{\futurelet\@let@token\@onedot}
\def\@onedot{\ifx\@let@token.\else.\null\fi\xspace}

\makeatother

\begin{document}
\let\WriteBookmarks\relax
\def\floatpagepagefraction{1}
\def\textpagefraction{.001}

\shorttitle{Artery-Vein Segmentation in DSA}

\shortauthors{Su et~al.}

\title [mode = title]{\method: Cerebral Artery-Vein Segmentation in Digital Subtraction Angiography}                      

\tnotetext[1]{This work was supported by Health-Holland (TKI Life Sciences and Health)through the Q-Maestro project under Grant EMCLSH19006 and Philips Healthcare (Best, The Netherlands).}


%
\author[1]{Ruisheng Su}[type=editor,
                        orcid=0000-0002-5013-1370]
\cormark[1]
\fnmark[1]
\ead{r.su@erasmusmc.nl}
\ead[url]{https://ruishengsu.github.io/}

\author[1]{P. Matthijs van der Sluijs}[]
\author[2]{Yuan Chen}[]
\author[1]{Sandra Cornelissen}
\author[1]{Ruben van den Broek}
\author[3]{Wim H. van Zwam}
\author[1]{Aad van der Lugt}
\author[1,4]{Wiro J. Niessen}
\author[5]{Danny Ruijters}
\author[1]{Theo van Walsum}

\affiliation[1]{organization={Biomedical Imaging Group Rotterdam, Department of Radiology~\&~Nuclear Medicine, Erasmus MC, University Medical Center Rotterdam, The Netherlands.},
    }
\affiliation[2]{organization={Department of Radiology~\&~Nuclear Medicine, UMass Chan Medical School, Worcester, USA.}}
\affiliation[3]{organization={Department of Radiology~\&~Nuclear Medicine, Maastricht UMC, Cardiovascular Research Institute Maastricht, The Netherlands.}}
\affiliation[4]{organization={Imaging Physics, Applied Sciences, Delft University of Technology, The Netherlands.}}
\affiliation[5]{organization={Philips Healthcare, Best, The Netherlands.}}

\cortext[cor1]{Corresponding author}



\begin{abstract}
Cerebral X-ray digital subtraction angiography (DSA) is a widely used imaging technique in patients with neurovascular disease, allowing for vessel and flow visualization with high spatio-temporal resolution. Automatic artery-vein segmentation in DSA plays a fundamental role in vascular analysis with quantitative biomarker extraction, facilitating a wide range of clinical applications. The widely adopted U-Net applied on static DSA frames often struggles with disentangling vessels from subtraction artifacts. Further, it falls short in effectively separating arteries and veins as it disregards the temporal perspectives inherent in DSA. To address these limitations, we propose to simultaneously leverage spatial vasculature and temporal cerebral flow characteristics to segment arteries and veins in DSA. The proposed network, coined \method{}, encodes a 2D+time DSA series using spatial modules, aggregates all the features using temporal modules, and decodes it into 2D segmentation maps. On a large multi-center clinical dataset, \method{} achieves a vessel segmentation Dice of 0.84 ($\pm$0.04) and an artery-vein segmentation Dice of 0.79 ($\pm$0.06). \method{} surpasses traditional Frangi-based $k$-means clustering (P$<$0.001) and U-Net (P$<$0.001) by a significant margin, demonstrating the advantages of harvesting spatio-temporal features. This study represents the first investigation into automatic artery-vein segmentation in DSA using deep learning. The code is publicly available at \url{https://github.com/RuishengSu/CAVE_DSA}.\par
\end{abstract}




\maketitle
\section{Introduction}
\subsection{Clinical background}
\begin{figure*}[!t]
\centering
\includegraphics[clip, trim=0cm 0cm 0cm 0cm, width=\textwidth]{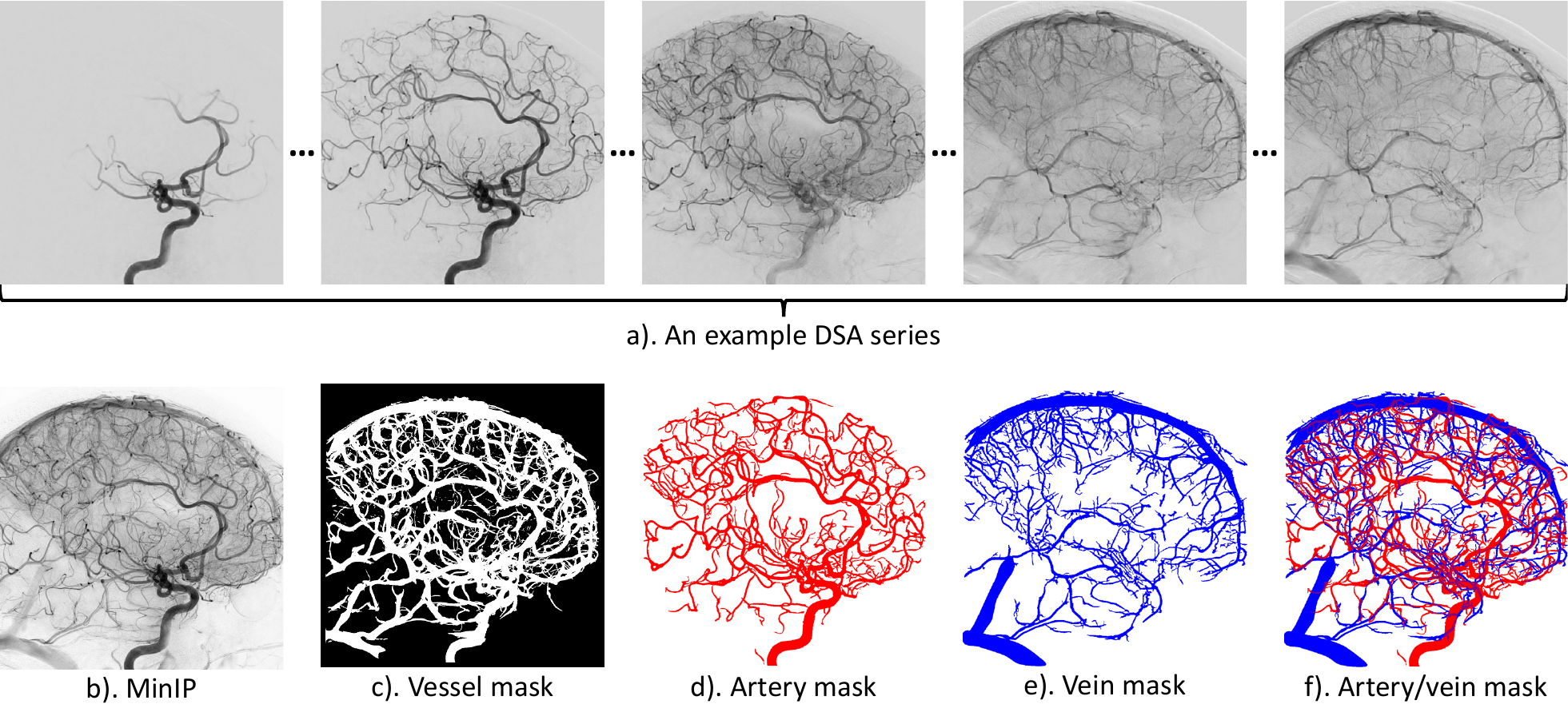}
\caption{Task illustration of artery-vein segmentation in DSA.} \label{fig:problem_definition}
\end{figure*}

Cerebrovascular diseases are a major contributor to global mortality and long-term disability~\citep{roth2020global}. These diseases encompass a range of conditions, including ischemic stroke due to vessel occlusion, stenosis, and aneurysms. In order to diagnose and treat such conditions, dynamic imaging of cerebral blood vessels is conducted using X-ray digital subtraction angiography (DSA). DSA provides a means of visualizing blood flow dynamics and changes in vasculature appearance over time (Figure~\ref{fig:problem_definition}), thereby offering valuable information for diagnosis, procedural navigation, therapeutic decision-making, and evaluation of treatment outcomes.\par

DSA images are conventionally examined visually by neuroradiologists and interventionalists, which could be laborious, subjective, qualitative, and vulnerable to error. Automatic segmentation of arteries and veins promises to assist in this assessment by highlighting and quantifying vascular changes, providing a foundation for a range of downstream clinical applications, including quantitative evaluation of endovascular thrombectomy, automatic emboli detection, and image guidance for real-time endovascular navigation. For example, automatic artery-vein segmentation can be valuable for providing a venous roadmap for navigation during transvenous procedures, such as dural venous sinus stenting or transvenous embolization of dural arteriovenous fistula or arteriovenous malformations. The segmented arteries and veins provide a wealth of quantitative data that can be used to extract and analyze various blood flow-related biomarkers for peri-operative decision-making and post-operative prognosis.\par

DSA is a dynamic imaging technique that provides a visual representation of blood flow over time through a series of consecutive frames (Figure~\ref{fig:problem_definition}). Although existing semantic segmentation networks could be utilized for end-to-end artery-vein segmentation, most methods only consider static frames. When addressing artery-vein segmentation in DSA series, which are 2D+time image sequences, the temporal dimension is relevant. We hypothesize that effectively incorporating spatio-temporal flow dynamics is key for achieving accurate artery-vein segmentation in DSA. The aim of this work, therefore is to harness the high-resolution contrast flow dynamics of DSA for improved artery-vein segmentation through spatio-temporal learning techniques.\par

\subsection{Related work}
Vessel segmentation is an extensively studied field in medical imaging for over two decades, with recent advancements propelled by deep learning. Recent literature on vessel segmentation~\citep{moccia2018blood} primarily addresses retinal~\citep{fraz2012blood,chen2021retinal}, lung~\citep{1610745,van2013automated,tan2021automated}, and cardiovascular imaging~\citep{huang2023extraction,huang2023nag}. These studies demonstrate the adaptability of segmentation methods across various anatomical structures and imaging modalities. In brain imaging, various methods have been proposed to segment cerebral vessels on different image modalities such as MRA~\citep{phellan2017comparison,phellan2017automatic,robben2016simultaneous}, CT angiography (CTA)~\citep{fu2020rapid,su2020automatic}, and DSA~\citep{liu2018vessel,su2021autotici,vepa2022weakly,zhang2020neural}, each modality posing unique challenges and necessitating tailored approaches. The prevalent use of U-Net~\citep{ronneberger2015u} in these studies underscores its versatility. However, existing methods mainly rely on spatial vasculature features learned from static 2D DSA frames without considering the complete series. It has been shown that U-Net tends to generate false positives in the presence of subtraction artifacts that appear similar to blood vessels~\citep{zhang2020neural}.\par

Artery-vein segmentation has been primarily explored in fundus images~\citep{hemelings2019artery} and non-contrast lung CT images~\citep{qin2021learning} using U-Net. However, the segmentation of cerebral arteries and veins remains an under-explored area, with limited studies on 4D CTA~\citep{meijs2020cerebral}, 3D MRA images~\citep{hilbert2020brave}, and 3D-DSA~\citep{raz2021arterial} using 3D U-Net models. These methods predominantly leverage spatial features~\citep{hilbert2020brave} or manually crafted temporal parameters~\citep{meijs2020cerebral} to differentiate between arteries and veins. In a previous study \citep{van2022automatic}, we investigated the automatic classification of arteries and veins using $k$-means clustering, contingent on vessel enhancement and binarization via the Frangi filter. The quality of vessel segmentation through the Frangi filter depends on the choice of the binarization threshold, a parameter that exhibits high variability across different DSA images. Nevertheless, to the best of our knowledge, no fully automatic and robust artery-vein segmentation algorithm has been developed for DSA images yet.\par

\subsection{Contributions}
This work presents \method, the first automatic method for artery-vein segmentation in DSA using deep learning, establishing a new benchmark. \method{} takes a 2D+time video sequence with variable length as input and produces 2D artery-vein segmentations as output. We leverage U-Net for spatial vasculature representation and temporal modules to learn temporal cerebral contrast flow characteristics simultaneously. On a multi-center clinical dataset, we demonstrate the utility of deep learning in cerebral artery-vein segmentation. \method{} may facilitate fast, accurate, and objective interpretation of cerebral vasculature in DSA, thus assisting endovascular interventions in clinical practice.\par 

The remainder of this paper is organized as follows: 
Section~\ref{sec:methods} describes the proposed artery-vein segmentation method and Section~\ref{sec:data} details the experimental dataset and annotation process. The extensive experiments and results are presented in Section~\ref{sec:exp_and_results}, and further discussed in Section~\ref{sec:discussion}. Finally, Section~\ref{sec:conclusion} summarizes the conclusions of this study.\par

\section{Methods} \label{sec:methods}
A cerebral digital subtraction angiography (DSA) series consists of a sequence of X-ray images captured post-contrast media injection, subtracted from the initial frame (pre-contrast image). This sequence presents the arrival of contrast media and the cerebral blood flow dynamics over time. The primary objective of this work is to develop an automated method for segmenting arteries and veins from a given DSA series, obtaining 2D artery-vein label masks as shown in Figure~\ref{fig:problem_definition}. \par

\begin{figure*}[!t]
\centering
\includegraphics[clip, trim=0cm 0cm 0cm 0cm, width=\textwidth]{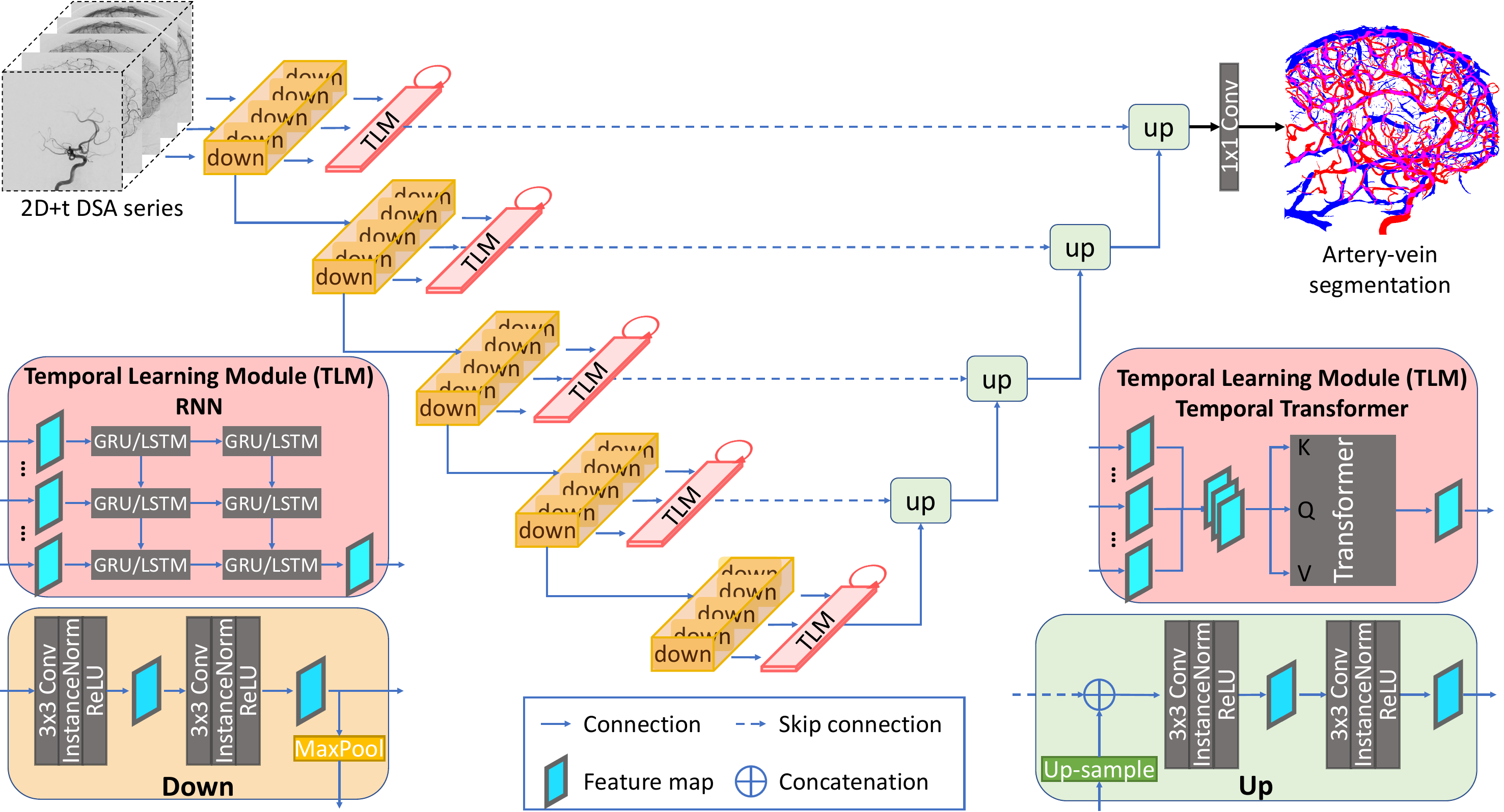}
\caption{Network architecture for artery-vein segmentation in DSA. Input is a 2D+time DSA series with variable series length. Output is a two-channel segmentation image with artery and vein represented by red and blue colors respectively. The temporal learning module (TLM) could be GRU, LSTM, or temporal transformer (TT).} \label{fig:architecture}
\end{figure*}

Rather than relying on pixel-wise time-intensity curves (TICs) or vasculature appearance separately, we propose to simultaneously leverage spatial and temporal features to segment arteries and veins in an end-to-end way. The proposed network architecture (Figure~\ref{fig:architecture}) takes DSA series of varying lengths as input. In the encoding path, each frame undergoes the same set of convolutional operations with shared weights, and the resulting features are temporally encoded to capture the spatio-temporal flow dynamics across all frames. These aggregated features are concatenated in the decoder path to generate high-resolution segmentation maps, resulting in a two-channel binary image that represents the segmented arteries and veins. The model is fully convolutional, thus allowing for input of varying sizes and temporal resolutions.\par

\subsection{Spatial learning}
The proposed \method{} (Figure~\ref{fig:architecture}) employs a UNet-like architecture for spatial encoding and decoding of DSA frames, which comprises five down layers (yellow boxes) and five up layers (green boxes). Each layer utilizes double-convolution blocks with instance normalization and ReLU activation. Max pooling and bilinear upsampling are utilized for the contracting and expanding path respectively. The number of feature channels starts from 64 and doubles after each max pooling to offset spatial information loss and halves after each bilinear upsampling operation to reconstruct spatial resolution. In Figure~\ref{fig:architecture}, it is only for illustration purposes that the down blocks appear to be replicated. In practice, all frames of the DSA series share the same down block for feature extraction, thereby avoiding an exponential increase in training parameters. \par

\subsection{Temporal learning}
We extend our approach beyond spatial representation learning by incorporating temporal learning techniques to improve the accuracy of artery-vein segmentation in DSA. Specifically, we explore three state-of-the-art techniques: convolutional GRU (ConvGRU)~\citep{ballas2015delving}, convolutional LSTM (ConvLSTM)~\citep{shi2015convolutional}, and temporal transformer (TT) based on~\cite{vaswani2017attention}. The convGRU and convLSTM architectures are designed to address the vanishing gradient problem~\citep{hochreiter1998vanishing} in long sequences. The convGRU employs update and reset gates, while the convLSTM uses forget, input, and output gates for a similar purpose. On the other hand, the transformer model leverages its attention mechanism to focus selectively on various segments of the input sequence, which could be particularly beneficial for understanding temporal relationships in the application of artery-vein segmentation in DSA. These techniques allow for the processing of input series of varying lengths and encoding them into fixed-sized representations, thereby enabling temporal aggregation of information. We apply this feature extraction and temporal aggregation process in all five layers, which operate on multi-scale image features.\par

Recurrent neural networks (RNN), such as GRU and LSTM, have been widely used in learning sequential information flow using a gate mechanism. In this work, we tailor ConvGRU and ConvLSTM to distinguish vessels from subtraction artifacts. GRU and LSTM models are employed to process sequential frame data, capturing temporal dependencies. The high-level architecture is shown in Figure~\ref{fig:architecture}. The module takes the spatial features of size $\mathit{B\times T\times C\times H\times W}$ generated from the U-Net encoders of each frame as input, and produces aggregated feature maps of size $\mathit{B\times C\times H\times W}$ that align in dimensions with the corresponding decoder branch. $\mathit{B}$, $\mathit{T}$, $\mathit{C}$, $\mathit{H}$, $\mathit{W}$ represents the batch size, frame length, channel size, image height, and image width respectively. In this work, the ConvGRU and ConvLSTM modules each consist of two convolutional RNN layers with a kernel size of $3\times 3$ and with hidden dimensions equal to input dimensions. Such an end-to-end trained module is designed to learn complex spatio-temporal features. It is worth noting that this module allows a variable number of input frames.\par

Besides RNN, Transformer~\citep{vaswani2017attention} has been shown effective in numerous deep learning tasks including image segmentation. We developed a temporal transformer module, as sketched in Figure~\ref{fig:architecture}, to learn aggregate temporal features through temporal attention. The temporal transformer module, with its temporal attention capability, is utilized to selectively emphasize pertinent frames in a DSA series to enhance the differentiation between arteries, veins, and subtraction artifacts based on temporal flow dynamics in DSA. The module takes as input the UNet-encoded feature maps of all frames, and aggregates temporal features along the time dimension with a kernel of $\mathit{1\times 1\times T}$. The resulting feature map maintains the same size as a single-frame feature map, which is then concatenated with the decoding branch to construct the high-resolution segmentation map. In this work, the temporal transformer module consists of one multi-head attention layer with a kernel size of $1\times 1$ and with hidden dimensions equal to input dimensions. \par

\subsection{Loss function} 
As suggested in \cite{isensee2021nnu}, we define the loss function for vessel segmentation ($\mathcal{L}_{vessel}(p, g)$) as a combination of a cross-entropy loss and a Dice loss~\citep{kervadec2023dice}: 
{\setlength{\mathindent}{0cm}
\begin{align}
\mathcal{L}_{vessel}(p, g) & = \mathcal{L}_{\mathrm{CE}}(p, g) + \mathcal{L}_{\mathrm{DSC}}(p, g),
\label{eq:loss_vessel}
\end{align}}where $p$ and $g$ denote the predicted class probability and the reference binary label respectively. For artery-vein segmentation, we similarly define the loss function $\mathcal{L}_{av}(p, g)$ as

{\setlength{\mathindent}{0cm}
\begin{align}
\begin{split}
\mathcal{L}_{av}(p, g) = & \mathcal{L}_{\mathrm{CE}}(p_a, g_a) + \mathcal{L}_{\mathrm{CE}}(p_v, g_v) \\
& + \mathcal{L}_{\mathrm{DSC}}(p_a, g_a) + \mathcal{L}_{\mathrm{DSC}}(p_v, g_v).
\label{eq:loss_av}
\end{split}
\end{align}}The subscript "$a$" and "$v$" denote artery and vein respectively. More specifically, the cross-entropy loss $\mathcal{L}_{CE}(p, g)$, defined as
{\setlength{\mathindent}{0cm}
\begin{align}
\mathcal{L}_{\mathrm{CE}}\left(p, g\right) = -g \log (p) + (1-g) \log (1-p),
\label{eq:ce_loss}
\end{align}}measures the difference between the reference labels and predicted class probabilities. The Dice loss $\mathcal{L}_{DSC}(p_a, g_a)$ and $\mathcal{L}_{DSC}(p_v, g_v)$ are defined as 

{\setlength{\mathindent}{0cm}
\begin{align}
\mathcal{L}_{\mathrm{DSC}}\left(p, g\right)= 1-\frac{2 \sum\limits_{i \in \Omega} g^{(i)} p^{(i)}}{\sum\limits_{i \in \Omega}\left[g^{(i)} + p^{(i)}\right]},
\label{eq:dice_loss}
\end{align}}where $\Omega$ denotes the subset of the image space where $g$ is positive.\par

\section{Data} \label{sec:data}
The study uses data from the MR CLEAN Registry \citep{jansen2018endovascular}. It is an observational cohort study that included patients with acute ischemic stroke from sixteen centers in the Netherlands between March 2014 and December 2018. From this multi-center clinical registry, we manually segmented arteries and veins on 97 DSA series from different patients with either anterior-posterior (AP) or lateral views. The DSA series were acquired using various imaging systems (e.g., Philips, GE, and Siemens). The size of the individual frames of the acquired DSA series is $1024\times1024$ pixels. The series have different lengths, ranging from 10 to 50 frames, and varying temporal resolutions ranging from 0.5 to 4 frames per second.\par


The artery-vein annotations were created using an in-house developed tool in MeVisLab~\citep{heckel2009object} by four trained clinical students. To reduce inter-observer variability, the annotations were further refined by another trained student. An experienced radiologist was available for consultation during annotation. As visualized in Figure~\ref{fig:problem_definition}, the annotation results in two segmentation images, with arterial annotation in red and venous annotation in blue. Overlapping pixels have both labels as the two annotations are independent. These annotations serve as the reference standard in this work.\par 

\section{Experiments and Results}\label{sec:exp_and_results}
\subsection{Implementation details} 
The proposed methods were implemented in PyTorch \citep{paszke2019pytorch}, all trained on an NVIDIA RTX A40 GPU. As pre-processing, we resized all frames to $512\times512$ pixels, linearly resampled the temporal resolution to 1 fps, and normalized the intensity values to $[0, 255]$. The corresponding mask images were also resized to the same resolution. In addition, data augmentation techniques, i.e., horizontal flipping, translation $\in [-5\%, 5\%]$ range, scaling $\in [-5\%, 5\%]$ range, and rotation $\in [-10^{\circ}, 10^{\circ}]$ range, were randomly applied during training with a probability of 0.5 for each.\par

We randomly split the dataset into training, validation, and testing set with a ratio of 50\%-20\%-30\% on the patient level. This resulted in 52 DSA series for training, 19 for validation, and 26 for testing. The models were trained using RMSprop optimization~\citep{hinton2012neural} and a ReduceLROnPlateau~\citep{paszke2019pytorch} scheduler with a patience of 10 epochs, a decay factor of 0.5, and an initial learning rate of \num{1e-5}. An early stopping strategy was applied with a patience of 50 epochs and a maximum of 1000 epochs.\par

\subsection{Baselines}
To benchmark the performance of \method{} and comprehensively assess the added value of simultaneous spatio-temporal learning in distinguishing vessels from subtraction artifacts, we implemented three representative solutions, and evaluated the \method{} against those.\par

\subpara{Frangi+$k$-means} 
We implemented a conventional two-step artery-vein classification method \citep{van2022automatic} that combines the Frangi filter and $k$-means clustering. First, the Frangi filter is applied to the static minimum intensity map (MinIP) of an input DSA, followed by fixed thresholding to obtain a binary vessel mask. Subsequently, all the vessel pixels within the mask are classified into arteries or veins. This clustering is achieved through $k$-means clustering with $k = 2$ using a set of intensity-based features derived from the time-intensity curve (TIC) of each pixel. These features include the area under the TIC curve (AUC), peak intensity, time to peak (TTP), arrival time, peak width, variance, standard deviation, and maximum uptake slope of the TIC. To ensure uniformity in the clustering process and mitigate the influence of varying magnitudes, all features are normalized to a range ranging from 0 to 1. Finally, the cluster with the lowest average TTP is identified as arterial, while the one with the highest average TTP is designated as venous. \par

\subpara{U-Net semantic segmentation} 
Apart from classical machine learning, a standard U-Net could be directly utilized for identifying arteries and veins leveraging the spatial vascular features. We implemented a deep learning baseline using U-Net~\citep{ronneberger2015u,zhang2020neural}. The architecture is similar to the architecture in Figure~\ref{fig:architecture} with the contracting path replaced by stand-alone down blocks. To have all vessels present in the spatial dimensions, this U-Net approach uses the MinIP image of all frames of a DSA series as input. Consequently, U-Net purely relies on spatial vasculature features to differentiate between arteries and veins without considering the temporal contrast flow features.\par

\subpara{U-Net+$k$-means} 
To utilize both spatial and temporal features for artery-vein segmentation, we further developed a two-stage baseline approach that cascades U-Net and $k$-means clustering that sequentially leverages spatial and temporal information. The U-Net encodes a MinIP image and produces a binary vessel segmentation. The vessel pixels are subsequently clustered into arteries and veins via $k$-means clustering on the time-intensity curves.\par

\subsection{Evaluation metrics}
We assess these methods on a hold-out test set composed of 26 DSA sequences from various patients from two perspectives: vessel segmentation and artery-vein segmentation (Table~\ref{tab:performance}). We report the Dice coefficients, accuracy, sensitivity, and specificity for vessel segmentation, respectively defined below. 
\begin{align}
\mathit{Dice} = \frac{\mathit{2\ TP}}{\mathit{2\  TP + FP + FN}}\quad, \label{eq:vessel_dice}
\end{align}
\begin{align}
\mathit{Acc} = \frac{\mathit{TP + TN}}{\mathit{TP + TN + FP + FN}}\quad, \label{eq:vessel_accuracy}
\end{align}
\begin{align}
\mathit{Sens} = \frac{\mathit{TP}}{\mathit{TP + FN}}\quad, \label{eq:vessel_sensitivity}
\end{align}
\begin{align}
\mathit{Spec} = \frac{\mathit{TN}}{\mathit{TN + FP}}\quad. \label{eq:vessel_specificity}
\end{align}

For artery-vein segmentation, we additionally evaluate the artery Dice ($\mathit{A\text{-}Dice}$), vein Dice ($\mathit{V\text{-}Dice}$), and multi-class Dice ($\mathit{M\text{-}Dice}$). We define $\mathit{A\text{-}Dice}$ and vein Dice $\mathit{V\text{-}Dice}$ as 
\begin{align}
\mathit{A\text{-}Dice} = \frac{\mathit{2\ TP_a}}{\mathit{2\  TP_a + FP_a + FN_a}}\quad, \label{eq:a_dice}
\end{align}
\noindent and
\begin{align}
\mathit{V\text{-}Dice} = \frac{\mathit{2\ TP_v}}{\mathit{2\ TP_v + FP_v + FN_v}}\quad,\label{eq:v_dice}
\end{align}
where the subscript $\mathit{a}$ and $\mathit{v}$ denote artery and vein respectively. TP, FP, and FN are true positive, false positive, and false negative respectively. We define the multi-class Dice ($\mathit{M\text{-}Dice}$) as

{\setlength{\mathindent}{0cm}
\begin{align}
\mathit{M\text{-}Dice} = \frac{\mathit{2\ (TP_a + TP_v)}}{\mathit{2\ (TP_a + TP_v) + FP_a + FP_v + FN_a + FN_v}}.
\label{eq:m_dice}
\end{align}}Besides, we compute statistical significance using the paired Wilcoxon test on the Dice coefficients.\par

\begin{table*}[!t]
\centering
\caption{Performance of \method{} and other existing methods in vessel and artery-vein segmentation on the test set. Acc: accuracy, Sens: sensitivity, Spec: specificity, A-Dice: Artery Dice (Eq. \ref{eq:a_dice}), V-Dice: Vein Dice (Eq. \ref{eq:v_dice}), M-Dice: Multi-class Dice (Eq. \ref{eq:m_dice}). The accuracy, sensitivity, and specificity of artery-vein segmentation are over both artery and vein classes.}\label{tab:performance}
\resizebox{\textwidth}{!}{
\begin{tabular}{lllllllllll}
\hline
\multirow{2}{*}{Method} & \multicolumn{4}{c}{Vessel segmentation} & \multicolumn{6}{c}{Artery-vein segmentation} \\ \cmidrule(lr){2-5} \cmidrule(l){6-11}
& Acc & Sens & Spec & \textbf{Dice} & Acc & Sens & Spec & A-Dice & V-Dice & \textbf{M-Dice}\\ \midrule

\multirow{2}{*}{\specialcell{Frangi + $k$-means}} & 0.83 & 0.69 & 0.89 & 0.70 & 0.82 & 0.63 & 0.89 & 0.65 & 0.54 & 0.60 \\

& $\pm$0.048 & $\pm$0.13 & $\pm$0.088 & $\pm$0.094 & $\pm$0.050 & $\pm$0.11 & $\pm$0.083 & $\pm$0.077 & $\pm$0.12 & $\pm$0.084 \\ [0.15cm]
 
\multirow[c]{2}{*}{\specialcell{U-Net}}   & 0.89 & 0.79 & 0.93 & 0.80 & 0.87 & 0.68& 0.94 & 0.70 & 0.63 & 0.67 \\ 
 & $\pm$0.032 & $\pm$0.071 & $\pm$0.029 & $\pm$0.050 & $\pm$0.041 & $\pm$0.078 & $\pm$0.028 & $\pm$0.082 & $\pm$0.070 & $\pm$0.060 \\ [0.15cm]
 
 \multirow[c]{2}{*}{\specialcell{U-Net + $k$-means}}   & 0.89 & 0.79 & 0.93 & 0.80 & 0.87 & 0.73 & 0.93 & 0.73 & 0.65 & 0.69 \\ 
 & $\pm$0.032 & $\pm$0.071 & $\pm$0.029 & $\pm$0.050 & $\pm$0.035 & $\pm$0.071 & $\pm$0.028 & $\pm$0.060 & $\pm$0.067 & $\pm$0.050 \\ [0.2cm] \midrule

 \multirow[c]{2}{*}{\specialcell{\method\\\textit{(Temporal Transformer)}}} & 0.91 & 0.83 & 0.94 & 0.84 & 0.90 & 0.79 & 0.95 & 0.82 & 0.74 & 0.78 \\
 & $\pm$0.029 & $\pm$0.064 & $\pm$0.025 & $\pm$0.041 & $\pm$0.032 & $\pm$0.071 & $\pm$0.024 & $\pm$0.054 & $\pm$0.071 & $\pm$0.048 \\ [0.15cm]

 \multirow[c]{2}{*}{\specialcell{\method\\\textit{(ConvLSTM)}}}   & 0.91 & 0.83 & 0.94 & 0.83 & 0.90 & 0.80 & 0.95 & 0.82 & 0.75 & 0.79 \\
 & $\pm$0.025 & $\pm$0.060 & $\pm$0.025 & $\pm$0.048 & $\pm$0.031 & $\pm$0.075 & $\pm$0.022 & $\pm$0.062 & $\pm$0.058 & $\pm$0.052 \\ 
 [0.15cm]
 
\multirow[c]{2}{*}{\specialcell{\method\\\textit{(ConvGRU)}}}   & 0.91 & 0.84 & 0.94 & 0.84 & 0.90 & 0.79 & 0.95 & 0.82 & 0.76 & 0.79 \\
 & $\pm$0.027 & $\pm$0.062 & $\pm$0.022 & $\pm$0.039 & $\pm$0.035 & $\pm$0.078 & $\pm$0.024 & $\pm$0.077 & $\pm$0.054 & $\pm$0.057 \\ [0.2cm] \hline
\end{tabular}}
\end{table*}

\begin{figure*}[!t]
\centering
\includegraphics[clip, trim=0cm 0cm 0cm 0cm, width=\textwidth]{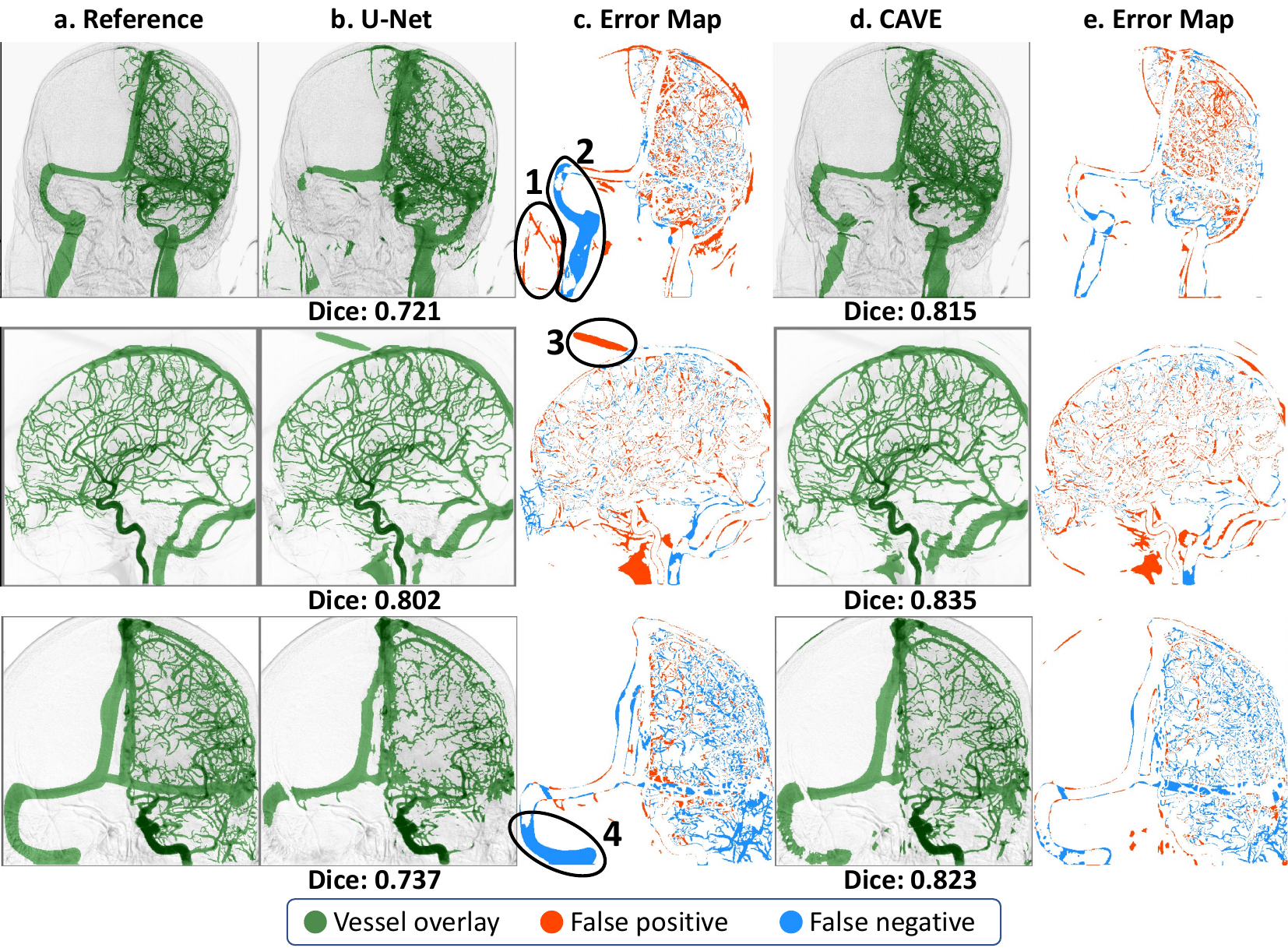}
\caption{Example visualizations of cerebral vessel segmentation results of U-Net and the proposed \method. Column a: manual annotation of vessels overlaid on the MinIP image; column b: segmentation output of U-Net; column c: U-Net error map with false positives (orange) and false negatives(light blue); column d: segmentation output of \method; column e: \method{} error map with false positives (orange) and false negatives (light blue).} \label{fig:qualitative_vessel}
\end{figure*}

\subsection{Quantitative analysis} 
In Table~\ref{tab:performance}, we present the results of \method{} and other representative solutions on the test set in terms of vessel segmentation and artery-vein segmentation. Regarding vessel segmentation, both U-Net and \method{} significantly outperform the Frangi+$k$-means approach in terms of Dice coefficient with a margin of 10\% (P$<$0.001). Notably, \method{} substantially surpasses U-Net by 3.6\% (P=0.023) and the Frangi+$k$-means approach by 14\% (P$<$0.001), demonstrating the effectiveness of incorporating simultaneous spatio-temporal vascular and flow representation for separating vessels from subtraction artifacts or other static instruments.\par

With respect to artery-vein segmentation, the advantage of spatio-temporal learning is even more prominent as shown in Table~\ref{tab:performance}. While Frangi+$k$-means, by relying solely on pixel-wise temporal characteristics, achieves a multi-class Dice coefficient of 0.60 ($\pm$0.084), spatial feature-based deep learning (U-Net) obtains a higher Dice coefficient of 0.67 ($\pm$0.060). The U-Net+$k$-means approach, which leverages spatial and temporal features in two sequential stages, achieves a Dice coefficient of 0.69 ($\pm$0.050). In contrast, by integrating both vasculature appearance and flow dynamics in an end-to-end spatio-temporal deep learning model, \method{} yields a significantly higher multi-class Dice coefficient of 0.79 ($\pm$0.057) compared to U-Net (P$<$0.001) and Frangi+$k$-means (P$<$0.001). Overall, \method{} demonstrates significant improvements in artery-vein segmentation by incorporating the temporal contrast flow aspects of DSA in end-to-end learning. We observe no statistically significant differences among ConvGRU, ConvLSTM, and the temporal transformer. \par

In terms of computational efficiency, \method{} takes on average 0.57$\pm$0.2 seconds to segment a complete DSA series on an NVIDIA 2080 Ti GPU with 11 GB of memory. \par

\begin{figure*}[!t]
\centering
\includegraphics[clip, trim=0cm 0cm 0cm 0cm, width=\textwidth]{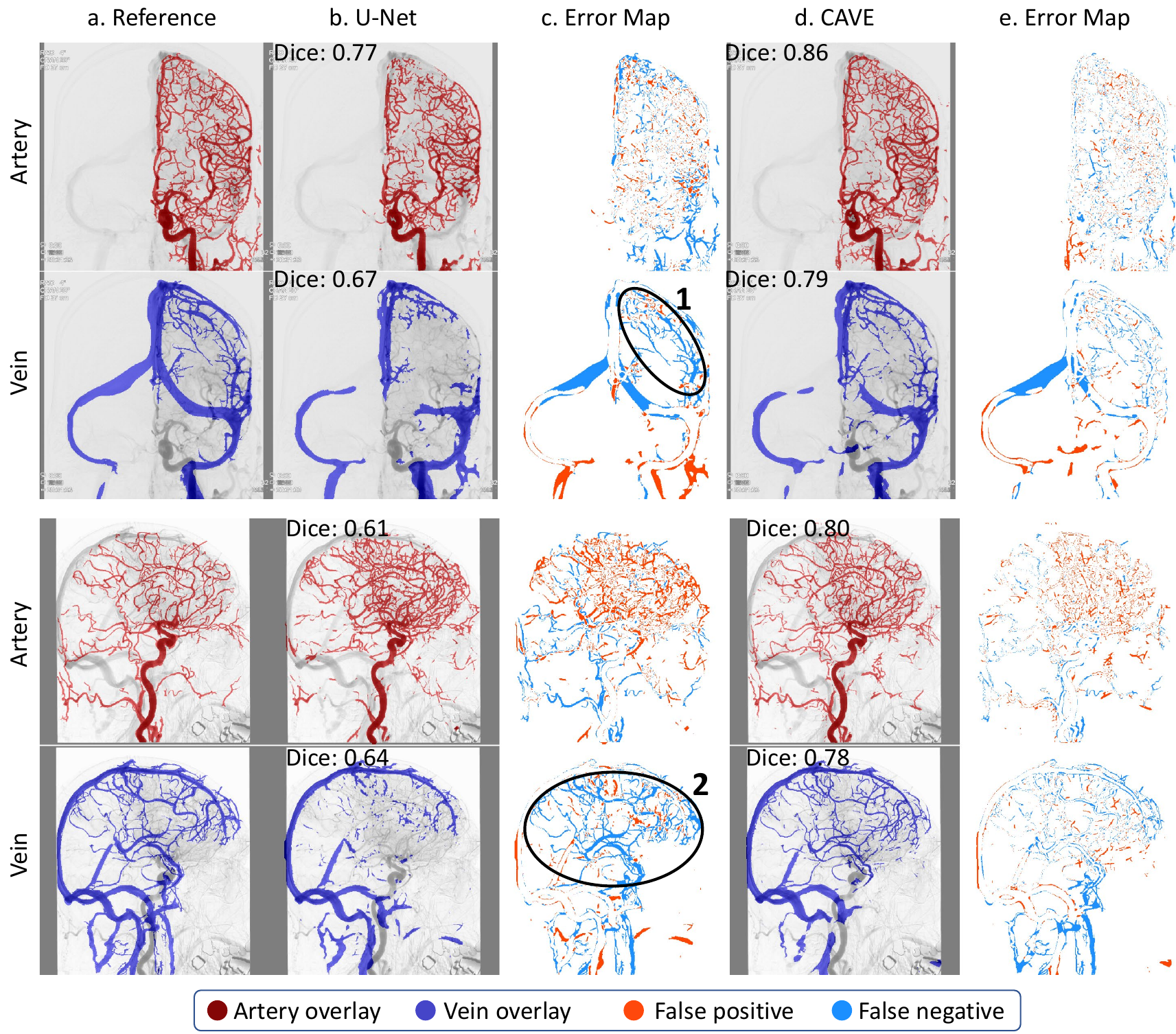}
\caption{Two example visualizations (in rows) of artery-vein segmentation. Column a: manual annotation of arteries (red) and veins (blue) overlaid on the MinIP image; column b: segmentation output of U-Net; column c: U-Net error map with false positives (orange) and false negatives(light blue); column d: segmentation output of \method; column e: \method{} error map with false positives (orange) and false negatives (light blue).} \label{fig:qualitative_av}
\end{figure*}

\subsection{Impact of temporal resolution}
\begin{figure}[!ht]
\centering
\includegraphics[clip, trim=0cm 0cm 0cm 0cm, width=\columnwidth]{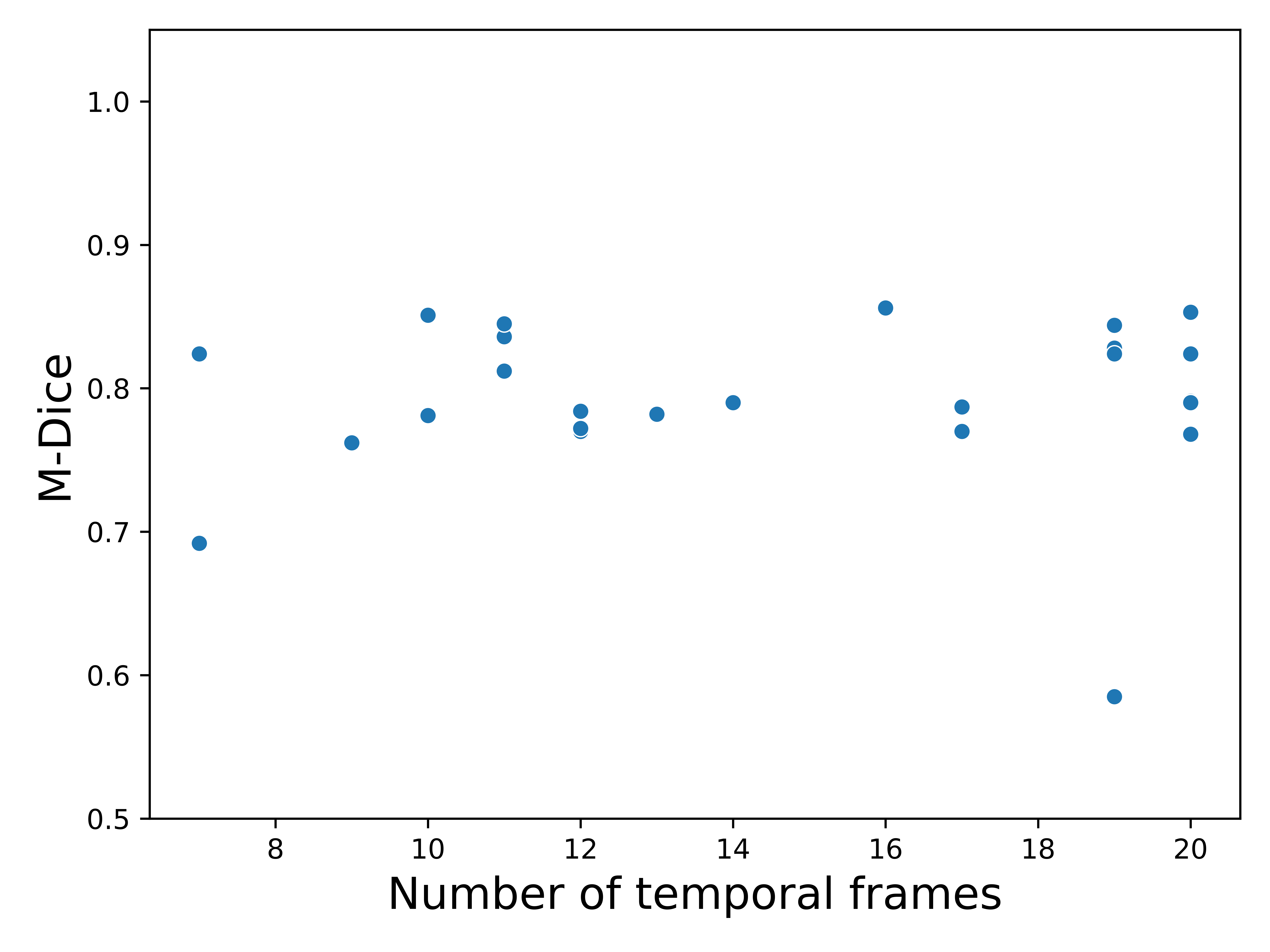}
\caption{Association between the segmentation Dice and the number of frames of DSA series in the test set.} \label{fig:number_of_frames}
\end{figure}

We investigate the influence of temporal characteristics, specifically the number of frames of a DSA series and the temporal frequency, on artery-vein segmentation performance. As shown in Figure~\ref{fig:number_of_frames}, we do not find a notable correlation between the average Dice coefficient and the frame count of DSA series in the test set, evidenced by a Pearson correlation coefficient of 0.05 (P=0.81). This could be attributed to the fact that all DSA series have the same temporal frequency and are complete in our dataset. In contrast, we observe a notable reduction in segmentation performance when the DSA series are temporally down-sampled for testing, as shown in Table~\ref{tab:DSA_frequency}. This suggests that the proposed \method{} method effectively utilizes temporal flow characteristics to enhance segmentation performance.

\begin{table}[!ht]
\centering
\caption{Performance of \method{} with respect to various temporal resolutions in frames per second (fps).}\label{tab:DSA_frequency}
\begin{tabular}{llll}
\hline
fps & 1 & 0.5 & 0.25 \\ \midrule
M-Dice  & 0.79$\pm$0.056 & 0.76$\pm$0.060 & 0.64$\pm$0.12 \\ [0.15cm] \midrule
\end{tabular}
\end{table}

\subsection{Qualitative analysis}
To provide a comprehensive comparison between the proposed \method{} and U-Net, we present representative visualizations and qualitative evaluations. \method{} is particularly effective in distinguishing cerebral vessels from the subtraction artifacts, instruments, and subtraction artifacts, thanks to its ability to learn temporal contrast flow dynamics. In Figure~\ref{fig:qualitative_vessel}, we provide visual comparisons of vessel segmentation results using U-Net and \method. The error maps, shown in columns c and e, use orange to indicate false positives and light blue to indicate false negatives. We identify two scenarios. First, in regions \#1 and \#3, where U-Net mistakenly recognizes subtraction artifacts and static instruments as vessels (column c), \method{} correctly avoids such misclassifications (column e). Furthermore, \method{} successfully identifies venous vessels, as shown in regions \#2 and \#4, even when they are surrounded by subtraction artifacts based on their temporal characteristics, whereas U-Net fails to detect them in the presence of noise in the background.\par

Figure~\ref{fig:qualitative_av} presents two examples illustrating the performance of U-Net and \method{} in artery-vein segmentation. Compared to manual annotations (column a), \method{} (column e) shows fewer errors than U-Net (column c). While U-Net performs well in recognizing large vessels such as proximal arteries and the superior sagittal sinus (SSS), it struggles to correctly classify distal arteries and veins, as highlighted in regions \#1 and \#2. In contrast, \method{} achieves accurate classification in both small and large vessels, demonstrating its effectiveness in capturing spatial and temporal features. Additional visual comparisons of these methods are available in the appendix. \par


\section{Discussion} \label{sec:discussion}
In this work, we proposed a fully automated deep learning-based method for artery-vein segmentation in cerebral DSA images. Through quantitative and qualitative comparisons, we have demonstrated the added value of simultaneous spatio-temporal learning (\method{}) against spatial learning (U-Net), temporal learning ($k$-means), and sequential two-stage spatial and temporal learning (U-Net + $k$-means). \par

This application-tailored solution innovatively leverages the distinctive temporal contrast flow characteristics in DSA to precisely identify cerebral arteries and veins while alleviating misclassifications of subtraction artifacts and other surgical instruments. The primary focus of the analyses revolves around evaluating the inherent benefits of harnessing temporal information. For quantitative assessments, we employ the foundational U-Net as our benchmark. With the rapidly evolving landscape of deep network architectures, the U-Net may not be positioned among the most recent models in the current literature. Recent advancements in state-of-the-art networks, such as UNet++ \citep{zhou2018unet++}, PointRend~\citep{kirillov2020pointrend}, nnU-Net~\citep{isensee2021nnu}, TransUNet~\citep{chen2021transunet}, or CTransNet~\citep{wang2022uctransnet} may outperform U-Net, when leveraging purely spatial features from MinIP images. Nevertheless, adapting the baseline spatial learning module and capitalizing on the abundant temporal flow dynamics are orthogonal and mutually reinforcing efforts to improve performance. \par

\method{} is designed to take complete DSA series with variable lengths and output an overall 2D artery-vein segmentation map for an input DSA series. The explored TLM modules are flexible in terms of input dimensions. These modules are capable of handling DSA series with variable lengths, while alternative methods such as 3D U-Net, direct feature concatenating, or temporal convolutional network (TCN) \citep{lea2016temporal} typically cannot. It necessitates temporal sliding or temporal padding/cropping to utilize models that require fixed frame length. Such models would also require temporarily varying annotations to train.\par

In comparison with the hold-out strategy, K-fold cross-validation may provide a more robust assessment of the described methods in this manuscript. We tested the proposed CAVE (with ConvGRU) method on five random data splits and obtained M-Dice coefficients of 0.79 ($\pm$0.057), 0.79 ($\pm$0.052), 0.78 ($\pm$0.043), 0.78 ($\pm$0.049), and 0.78 ($\pm$0.057). No significant performance differences (P=0.4, Kruskal-Wallis test) were observed among these splits. This indicates that the hold-out method allows for reliable performance estimates. Considering this observation, alongside the substantial computational demands of cross-validation, we opted for the hold-out strategy in subsequent performance evaluations. \par

In this study, we observe that, despite the differences in network design, the TLM modules (i.e., ConvGRU, ConvLSTM, and Temporal Transformer) perform comparably in the application of artery-vein segmentation on our dataset. This could be attributed to their shared capability to capture temporal dependencies. While the temporal transformer is often considered superior~\citep{vaswani2017attention,devlin2018bert,dai2019transformer}, its effectiveness may not consistently extend to all scenarios. In this context, temporal information is relevant. All modules do the job of extracting temporal patterns, whereas the exact network design and configuration do not make a notable difference. This aligns with a broader understanding that the practical impact of architectural advantages can vary greatly depending on the specifics of the task and data. \par


We note that the segmentation Dice scores of veins are on average lower than those of arteries across all methods. This observation can be attributed to several factors: 1) venous vessels typically exhibit relatively lighter intensities and more ambiguous boundaries, making them challenging to be accurately predicted; 2) the quality of vein annotations may also be comparatively lower due to increased annotation difficulties; 3) there tends to be a scarcity of venous phase frames, especially in cases of incomplete DSA series when image acquisition is prematurely stopped by the operator.\par

Building upon this work, future research may investigate the influence of patient motion on those spatio-temporal techniques for artery-vein segmentation. From the clinical perspective, there is significant value in conducting thorough validations to assess the clinical generalizability of \method{}.\par


\section{Conclusion} \label{sec:conclusion}
We have presented \method, a deep learning-based cerebral vessel and artery-vein segmentation method in digital subtraction angiography. It integrates both spatial vascular appearance and temporal contrast flow dynamics in a unified end-to-end framework, producing high-quality multi-class segmentations from cerebral DSA series with variable lengths. Experimental results on a multi-center clinical dataset demonstrate that \method{} significantly outperforms existing methods. Qualitative analyses further underpin the advantages of \method{} in distinguishing vessels from subtraction artifacts. \method{} has the potential to facilitate vessel-based quantitative analyses for clinical diagnosis, prognosis, and treatment planning in endovascular interventions.\par

\section*{Acknowledgments}
We want to thank the MR CLEAN Registry investigators for their contributions. The MR CLEAN Registry was funded and carried out by the Erasmus University Medical Centre, Amsterdam University Medical Centers, location AMC, and Maastricht University Medical Centre. The study was additionally funded by the Applied Scientific Institute for Neuromodulation (Toegepast Wetenschappelijk Instituut voor Neuromodulatie).\par



\bibliographystyle{cas-model2-names}

\bibliography{cas-refs}





\clearpage
\appendix
\section{Extended qualitative comparisons}


\noindent
\begin{minipage}{\textwidth}
\centering
\includegraphics[width=\textwidth]{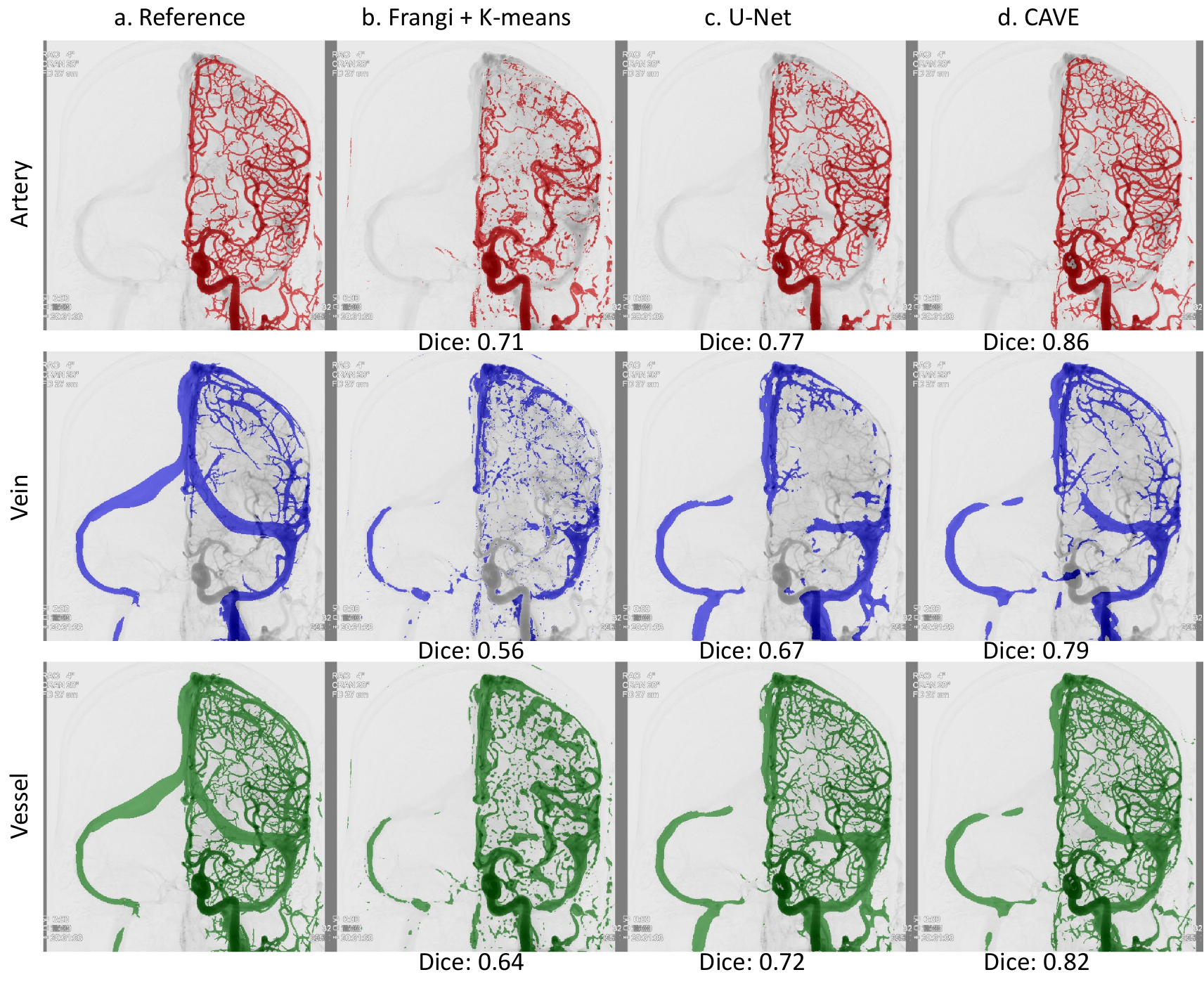}
\captionof{figure}[]{Example visualization of artery, vein, and vessel segmentation. Column a: manual annotation of arteries (red), veins (blue), and vessels (green) overlaid on the MinIP image; column b: segmentation output of the Frangi + $k$-means approach; column c: segmentation output of U-Net; column d: segmentation output of \method{}.} 
\end{minipage}

\clearpage
\noindent
\begin{minipage}{\textwidth}
\centering
\includegraphics[width=\textwidth]{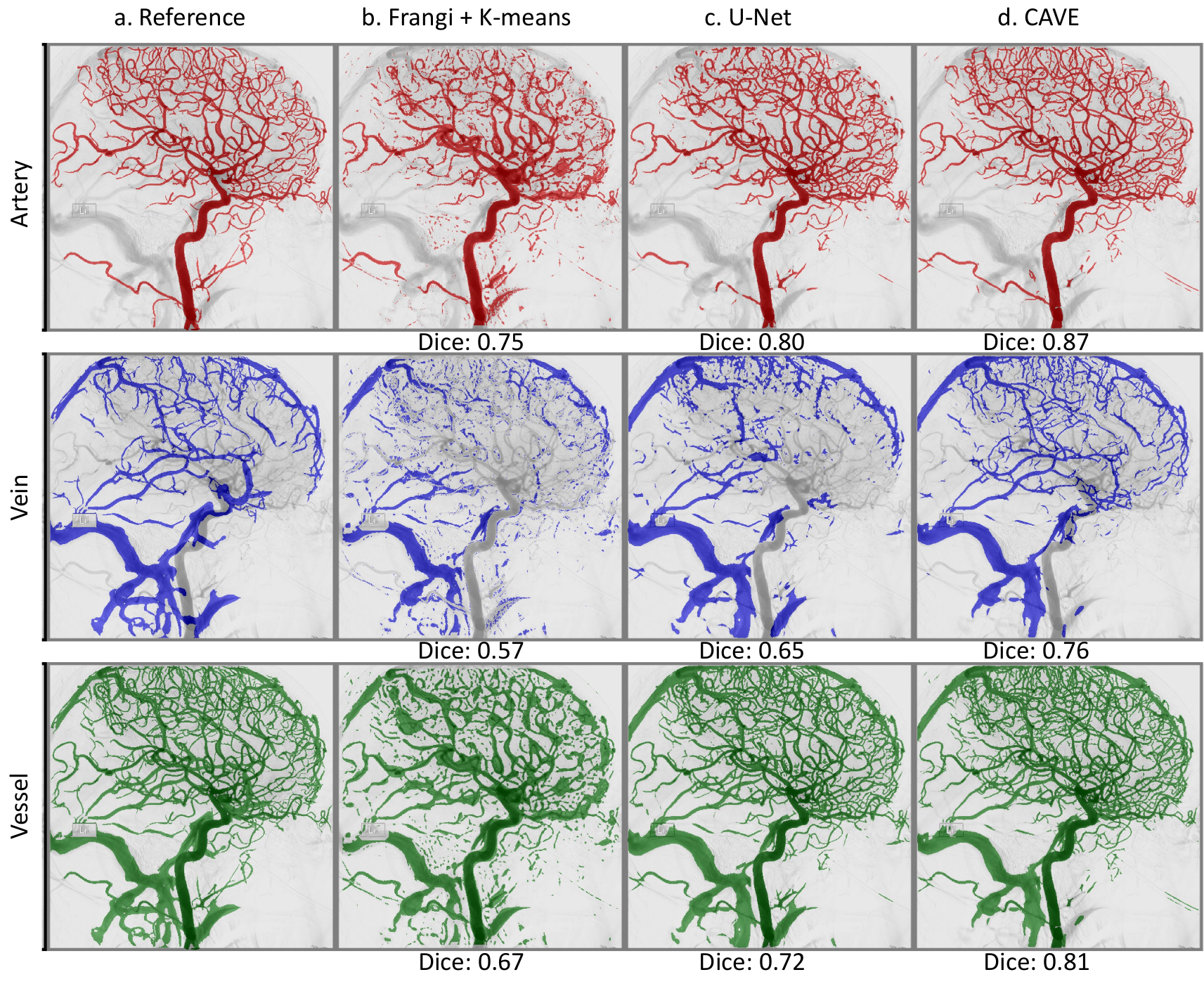}
\captionof{figure}[]{Example visualization of artery, vein, and vessel segmentation. Column a: manual annotation of arteries (red), veins (blue), and vessels (green) overlaid on the MinIP image; column b: segmentation output of the Frangi + $k$-means approach; column c: segmentation output of U-Net; column d: segmentation output of \method{}.} 
\end{minipage}

\end{document}